\documentclass[prd, aps, floatfix, twocolumn, superscriptaddress, preprintnumbers]{revtex4-1}

\usepackage[colorlinks = true, citecolor = blue, linkcolor = blue, urlcolor = blue]{hyperref}

\usepackage{graphicx}
\usepackage{tabularx}
\usepackage{makecell}
\usepackage[normalem]{ulem}

\usepackage{mathtools}
\usepackage{amsmath}
\usepackage[capitalise]{cleveref} 
\usepackage{amssymb}
\usepackage{bm}
\usepackage{mathrsfs}
\usepackage{slashed}
\usepackage{physics}
\usepackage[dvipsnames]{xcolor}
\usepackage{diagbox}
\allowdisplaybreaks

\usepackage{atbegshi, picture}
\AtBeginShipoutNext{\AtBeginShipoutUpperLeft{%
\put(\dimexpr\paperwidth-2.75cm\relax,-1.1cm){\makebox[0pt][r]{MIT-CTP/5668}}%
}}
\begin{document}
\title{Transverse Energy-Energy Correlator for Vector Boson-Tagged \\ Hadron Production in $pp$ and $pA$ collisions}

\author{Zhong-Bo Kang}
\email{zkang@physics.ucla.edu}
\affiliation{Department of Physics and Astronomy, University of California, Los Angeles, CA 90095, USA}
\affiliation{Mani L. Bhaumik Institute for Theoretical Physics, University of California, Los Angeles, CA 90095, USA}
\affiliation{Center for Frontiers in Nuclear Science, Stony Brook University, Stony Brook, NY 11794, USA}

\author{Sookhyun Lee}
\email{shlee@bnl.gov}
\affiliation{Physics Department, University of Tennessee, Knoxville, TN 37996, USA}
\affiliation{Physics Department, University of Michigan, Ann Arbor, MI 48109, USA}

\author{Jani Penttala}
\email{janipenttala@physics.ucla.edu}
\affiliation{Department of Physics and Astronomy, University of California, Los Angeles, CA 90095, USA}
\affiliation{Mani L. Bhaumik Institute for Theoretical Physics, University of California, Los Angeles, CA 90095, USA}

\author{Fanyi Zhao}
\email{fanyi@mit.edu}
\affiliation{Center for Theoretical Physics, Massachusetts Institute of Technology, Cambridge, MA 02139, USA}

\author{Yiyu Zhou}
\email{yiyu.zhou@unito.it}
\affiliation{Department of Physics, University of Turin, via Pietro Giuria 1, I-10125 Torino, Italy}
\affiliation{Key Laboratory of Atomic and Subatomic Structure and Quantum Control (MOE), Guangdong Basic Research Center of Excellence for Structure and Fundamental Interactions of Matter, Institute of Quantum Matter, South China Normal University, Guangzhou 510006, China}
\affiliation{Guangdong-Hong Kong Joint Laboratory of Quantum Matter, Guangdong Provincial Key Laboratory of Nuclear Science, Southern Nuclear Science Computing Center, South China Normal University, Guangzhou 510006, China}
\affiliation{Department of Physics and Astronomy, University of California, Los Angeles, CA 90095, USA}

\begin{abstract}

We investigate the transverse energy-energy correlator (TEEC) event-shape observable for back-to-back $\gamma + h$ and $Z + h$ production in both $pp$ and $pA$ collisions. Our study incorporates nuclear modifications into the transverse-momentum dependent (TMD) factorization framework, with resummation up to next-to-leading logarithmic (NLL) accuracy, for TEEC as a function of the variable $\tau = \pqty{1 + \cos{\phi}}/2$, where $\phi$ is the azimuthal angle between the vector boson and the final hadron. We analyze the nuclear modification factor $R_{pA}$ in $p\mathrm{Au}$ collisions at RHIC and $p\mathrm{Pb}$ collisions at the LHC. Our results demonstrate that the TEEC observable is a sensitive probe for nuclear modifications in TMD physics. Specifically, the changes in the $\tau$-distribution shape provide insights into transverse momentum broadening effects in large nuclei, while measurements at different rapidities allow us to explore nuclear modifications in the collinear component of the TMD parton distribution functions in nuclei.

\end{abstract}

\maketitle

\section{Introduction}

Event-shape observables are inherently sensitive to various energy scales anywhere between the hard scale and the non-perturbative scale of Quantum Chromodynamics (QCD).
Along with the collision system and the entities of choice that define an event-shape observable, how we define the observable can magnify particular aspects of the event topology, allowing us to zoom in on the underlying physics of interest.
At $e^+e^-$ and $ep$ colliders, event shapes traditionally have played a crucial role in determining the strong coupling constant $\alpha_s$~\cite{Abbate:2010vw,Hoang:2015hka,H1:2017bml}.
At the Large Hadron Collider (LHC), event-shape observables suitable for $pp$ collisions have also been introduced~\cite{Banfi:2010xy} and, in particular, observables that use jets as inputs in multi-jet events have been extensively measured in the past decade~\cite{CMS:2011usu, ATLAS:2012tch, CMS:2013lua, CMS:2014tkl, ATLAS:2015yaa, ATLAS:2017qir, CMS:2018svp, ATLAS:2023tgo}.
These observables are known for their high precision in theory calculations \cite{Czakon:2021mjy, Alvarez:2023fhi}.
Recently, observables that define their shapes in terms of particles inside jets rather than jets themselves are often measured \cite{ATLAS:2012uka,ATLAS:2019rqw,ALICE:2019ykw,ALICE:2021njq,ALICE:2021vrw}.
Such observables provide enhanced sensitivities to collinear and soft emissions.
For this reason, their measurements allow for in-depth studies of radiation patterns and non-perturbative effects and are used to fine-tune parton shower and hadronization models in Monte Carlo simulations. 
Furthermore, fully global event shapes \cite{Kang:2013nha,Kang:2013lga,Cao:2024ota} have been computed with a state-of-the-art theoretical precision for deep inelastic scattering (DIS), where experimentally clean environment is advantageous in achieving high-precision measurements. 
These types of observables are gaining increasing attention due to their potentially greater handle on non-perturbative QCD power corrections. Some have been measured recently \cite{H1:2024aze,H1:2024pvu}, and theoretical development is ongoing for measurements planned at the future Electron-Ion Collider (EIC) \cite{Boer:2011fh, Accardi:2012qut, AbdulKhalek:2021gbh, AbdulKhalek:2022hcn}.
Additionally, event-shape observables are great tools for discovering new physics phenomena, \textit{e.g.}, constraining new colored matter \cite{Kaplan:2008pt, Llorente:2018wup}.

In this paper, we will focus in particular on the transverse energy-energy correlator (TEEC) event-shape observable.
TEEC~\cite{Ali:1984yp} is an extension of the energy-energy correlator (EEC)~\cite{Basham:1978bw, Basham:1978zq}, which was initially introduced for $e^+e^-$ collisions to characterize global event shapes, and has been broadly investigated at different experiments \cite{SLD:1994idb, L3:1992btq, OPAL:1991uui, TOPAZ:1989yod, TASSO:1987mcs, JADE:1984taa, Fernandez:1984db, Wood:1987uf, CELLO:1982rca, PLUTO:1985yzc, OPAL:1990reb, ALEPH:1990vew, L3:1991qlf, SLD:1994yoe}.
Later on, TEEC was defined such that it incorporates the transverse energy of the hadrons and is a more suitable observable for hadronic collider environments \cite{ATLAS:2015yaa, ATLAS:2017qir, ATLAS:2020mee, Ali:2012rn, Gao:2019ojf}.
Recently, TEEC has been computed for lepton-hadron production in lepton-proton scattering \cite{Li:2020bub, Kang:2023oqj}.
At the LHC, the EEC has also been measured using particles inside jets \cite{CMS:2024mlf, ALICE:2024dfl}.

A great advantage of the EEC and TEEC observables over other event-shape observables is that they effectively suppress contributions from soft radiation of low-energy nature and therefore become less sensitive to hadronization effects.
Another advantage of the TEEC lies in the accurate reconstruction of collision kinematics in the laboratory frame as highlighted in Ref. \cite{Gao:2022bzi}. 
In the back-to-back limit of the EEC and TEEC, results can be written directly in terms of the transverse-momentum dependent (TMD) parton distribution functions, enabling highly accurate predictions upon the resummation of Sudakov logarithms.
All of these make the TEEC a unique lens through which to investigate the transverse-momentum dependent structures of the proton \cite{Li:2020bub, Kang:2023big} and advance our understanding of non-perturbative dynamics of QCD.

Measuring hadrons in jets recoiling against a vector boson, $Z$ or $\gamma$, allows us to focus on a certain parton flavor as has been exploited in studying jet fragmentation properties in $pp$ collisions \cite{ATLAS:2019dsv,LHCb:2019qoc, CMS:2021iwu, LHCb:2022rky}.
Utilizing all hadrons in the back-to-back limit of this process when measuring the TEEC observable provides a clean access to the gluon TMD and the non-perturbative component of the light-quark TMD fragmentation function (FF).
Experimentally, low-energy hadrons are challenging to measure because of a high level of background consisting of spurious particles and large measurement uncertainties.
The TEEC observables are defined in such a way that they naturally suppress soft particles, reducing the sensitivity to the low-energy background.
In addition, the absence of the jet radius parameter and collinear-soft functions simplifies theoretical computations.
The TEEC for $\gamma/Z$-jet production in hadron colliders has been recently studied in \cite{Gao:2023ivm} at next-to-next-to-next-to-leading logarithmic ($\mathrm{N}^3\mathrm{LL}$) accuracy.

Another aspect of this process that can be explored is its sensitivity to the nuclear medium \cite{Kartvelishvili:1995fr, Wang:1996pe, Wang:1996yh, Wang:2013cia, Dai:2012am, Kang:2017xnc, Qin:2009bk}.
As the vector boson does not experience strong interaction, its momentum is not modified by the medium. This enables us to quantify the transverse momentum of the parton produced in the initial-state scattering.
The jet, on the other hand, loses its energy through interactions with the medium.
Measuring how much energy is lost leads us to gain insights into the role that nuclear medium plays in altering the initial partonic dynamics inside the proton.
The nuclear modification of this kind has been observed in measurements of the dijet and photon-jet momentum imbalance \cite{ATLAS:2017xfa, CMS:2012ulu, ATLAS:2018dgb, CMS:2017ehl}, along with jet fragmentation functions \cite{ATLAS:2018bvp, CMS:2014jjt}.
It is therefore of great interests to study if similar effects can be found for the TEEC between vector bosons and hadrons.

The rest of the paper is organized as follows.
In \cref{s.theory}, we present the theoretical formalism for the TEEC in $pp$ to boson-hadron production.
With the factorization at hand, all the ingredients appearing in the factorization formula are discussed.
They include the hard function, TMD parton distribution functions (PDFs), TEEC jet functions and soft function.
Additionally, we provide the TMD PDFs and TEEC jet functions in a nucleus that are needed to study nuclear modification for the TEEC in $pA$ collisions.
In \cref{s.phenomenology}, we explore phenomenological impact of the TEEC as a potential probe into these nuclear TMD PDFs and TEEC jet functions at both RHIC and LHC kinematics. 
Finally, we summarize our work in \cref{s.conclusion}.

\section{Theoretical formalism}
\label{s.theory}

In this section, we outline the factorization of the transverse energy-energy correlation between a vector boson and final-state hadrons in the back-to-back limit.
The process is illustrated in \cref{f.reaction} and is given as:
\begin{figure*}[tb]
\centering
\includegraphics[width = 0.87 \textwidth]{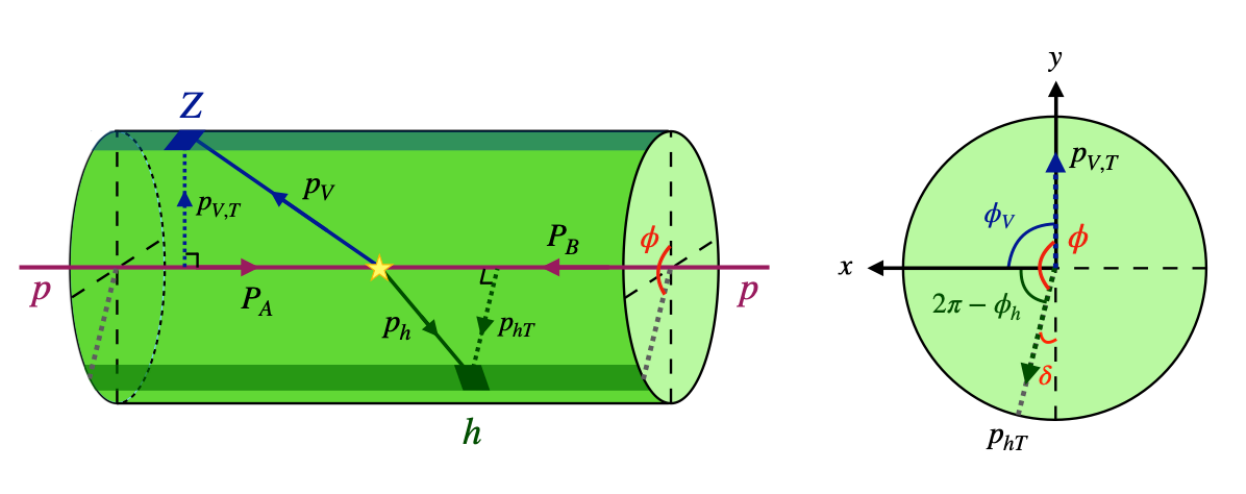}
\caption{Left: Illustration of TEEC in the $pp$ center-of-mass frame.
The incoming proton momentum $P_A$ and $P_B$ define the $z$-axis.
The transverse momentum of the outgoing vector boson $\boldsymbol{p}_{V,T}$ defines the $+y$-direction and hence the $xy$-plane is determined.
Right: The $xy$-plane of the left panel.}
\label{f.reaction}
\end{figure*}
\begin{align}
p \pqty{P_A} + p \pqty{P_B}
\to
V \pqty{p_V} + h \pqty{p_h} + X
\, , \label{e.hadronic-process}
\end{align}
where $p$, $V$ and $h$ denote the initial-state proton, final-state vector boson (either photon or $Z$-boson) and final-state hadron.
The momentum of each particle is given in parenthesis.
With $E_{T,V}$ and $E_{T,h}$ denoting the transverse energy of the vector boson $V$ and the final-state hadron $h$, we define the TEEC as:
\begin{align}
\textbf{TEEC}
& =
\sum_{h} \int \dd{\sigma}
\frac{E_{T,V} E_{T,h}}{E_{T,V} \sum_{h'} E_{T,h'}}
\delta \pqty{\tau-\frac{1+\cos{\phi}}{2}}
\nonumber \\
& =
\sum_{h} \int \dd{\sigma}
\frac{E_{T,h}}{\sum_{h'} E_{T,h'}}
\delta \pqty{\tau-\frac{1+\cos\phi}{2}}
\, .
\label{e.TEEC}
\end{align}
In this definition, the contributions from all the final-state hadrons are summed over. The variable $\tau$ is defined by $\tau \equiv \pqty{1 + \cos{\phi}}/2$,
where $\phi$ is the azimuthal angle between the vector boson and hadron $h$ in the $xy$-plane as illustrated in \cref{f.reaction}.
One can easily find that in the back-to-back configuration where $\phi$ approaches $\pi$, $\pi - \phi = \pi - (2 \pi - \phi_h + \phi_V) = \phi_h - \phi_V - \pi$ is a small angle, and $\tau \ll 1$.

The factorization formalism of the TEEC as defined in \cref{e.TEEC} is then expressed as:
\begin{widetext}
\begin{align}
\frac{\dd{\Sigma}}{\dd{\tau} \dd{y_V} \dd{p_{V,T}}}
& =
\sum_{a,b,c} \frac{p_{V,T}}{\sqrt{\tau}}
\int_{-\infty}^{\infty} \frac{\dd{b}}{2 \pi}
e^{-2i b \sqrt{\tau} p_{V,T}}
\int \dd{y_c}
H_{ab \to Vc} \pqty{y_V, p_{V,T}, m_V, \mu}
\nonumber \\
& \qquad \qquad \qquad \qquad \times
x_a \, f_{1,a/p}^{(u)} \pqty{x_a, b, \mu, \frac{\zeta_a}{\nu^2}}
x_b \, f_{1,b/p}^{(u)} \pqty{x_b, b, \mu, \frac{\zeta_b}{\nu^2}}
J_c^{(u)} \pqty{b, \mu, \frac{\zeta_c}{\nu^2}}
S_{abc} \pqty{b, \mu, \nu}
\nonumber \\
& =
\sum_{a,b,c} \frac{p_{V,T}}{\sqrt{\tau}}
\int_0^{\infty} \frac{\dd{b}}{\pi}
\cos(2 b  \sqrt{\tau} p_{V,T})
\int \dd{y_c}
H_{ab \to Vc} \pqty{y_V, p_{V,T}, m_V, \mu}
\nonumber \\
& \qquad \qquad \qquad \qquad \times
x_a \, f_{1,a/p}^{(u)} \pqty{x_a, b, \mu, \frac{\zeta_a}{\nu^2}}
x_b \, f_{1,b/p}^{(u)} \pqty{x_b, b, \mu, \frac{\zeta_b}{\nu^2}}
J_c^{(u)} \pqty{b, \mu, \frac{\zeta_c}{\nu^2}}
S_{abc} \pqty{b, \mu, \nu}
\, , \label{e.unsubtracted-factorization}
\end{align}
\end{widetext}
where $y_V$ and $p_{V,T}$ denote the rapidity and transverse momentum of the vector boson, respectively.
The subscripts $a$, $b$ and $c$ in $H_{ab \to Vc}$ can take a parton species, $q$ for a quark, $\bar q$ for an antiquark, and $g$ for a gluon.
Here $y_c$ is the rapidity of the parton $c$ that initiates the TEEC jet $J_c$, and $m_V$ is the mass of the produced vector boson.
The $y_c$ dependence is implicit and will be detailed in \cref{ss.hard-function-and-TMD-PDFs}.

Since the momentum imbalance is along the $x$ axis by definition, we have $\boldsymbol{b} \equiv \pqty{b_x, b_y} = \pqty{b, 0}$.
As a result, the dimensionality of Fourier transform becomes 1 as previously pointed out in, \textit{e.g.}, \cite{Li:2020bub, Fang:2023thw, Gao:2023ulg}.
In the factorization formalism, $f_{1,i/p}^{(u)} \pqty{x_i, b, \mu, \zeta_i/\nu^2}$ for parton species $i$ is its ``unsubtracted'' TMD PDF, where $x_i$ is the collinear momentum fraction of the parent hadron carried by parton $i$ and $\zeta_i$ is the Collins-Soper parameter.
It is worth noting that $b$ contains only the $x$-component as discussed above. 
The ``unsubtracted'' TMD PDFs describe energetic radiations of incoming partons in the direction collinear to the incoming protons \cite{Boussarie:2023izj}.

The hard function $H_{ab \to Vc} \pqty{y_V, p_{V,T}, m_V, \mu}$ contains the virtual corrections to the underlying partonic process $ab \to Vc$ that can be obtained through matching calculations between QCD and the soft-collinear effective theory (SCET).
The wide-angle soft emissions connecting initial and final colored particles $abc$ are encoded in the soft function $S_{abc} \pqty{b, \mu, \nu}$.
Finally, $J_c^{(u)} \pqty{b, \mu, \zeta_c/\nu^2}$ is the ``unsubtracted'' TEEC jet function, which has a close relation with the TMD fragmentation functions, and will be discussed in \cref{ss.TEEC}.
To arrive at the last line of \cref{e.unsubtracted-factorization}, we exploited the fact that $f_{1,i/p}^{(u)}$, $S_{abc}$ and $J_c^{(u)}$ are all even functions of $b$ (they all depend on $b^2$).
The TMD PDFs, soft function and TEEC jet function also depend on the renormalization scale $\mu$ and the rapidity renormalization scale $\nu$ \cite{Chiu:2012ir}.
Furthermore, $\zeta_{a, b, c}$ in TMD PDFs or TEEC jet function are the Collins-Soper parameters and will be discussed in \cref{ss.hard-function-and-TMD-PDFs,ss.TEEC}.

In the rest of this section, we will quantitatively define and discuss in details each ingredient in the factorization formula in \cref{e.unsubtracted-factorization}.

\subsection{Hard function and TMD PDFs}
\label{ss.hard-function-and-TMD-PDFs}

In \cref{e.unsubtracted-factorization}, $H_{ab \rightarrow  Vc}\pqty{y_V, p_{V,T},\mu}$ is the hard function that describes the partonic scattering process $ab \rightarrow Vc$.
The hard function $H_{ab \rightarrow Vc}$ at LO is given as \cite{Chien:2019gyf, Chien:2022wiq}:
\begin{align}
H_{q \overline{q} \to Vg}^{(0)}
& =
\frac{\mathcal{N}_H \pqty{N_c^2-1}}{N_c^2}
\frac{\hat{t}^2 + \hat{u}^2 + 2 \hat{s} m_V^2}{\hat{t} \hat{u}}
\, , \label{e.hard-qqb} \\
H_{q g \to Vq}^{(0)}
& =
-\frac{\mathcal{N}_H}{N_c}
\frac{\hat{s}^2 + \hat{t}^2 + 2 \hat{u} m_V^2}{\hat{s} \hat{t}}
\, , \label{e.hard-qg}
\end{align}
where $N_c$ is number of colors for quarks and the normalization $\mathcal{N}_H$ given by:
\begin{align}
\mathcal{N}_H
\equiv
\frac{2\pi p_{V,T}}{ \hat{s}^2}
\alpha_s \alpha_{\mathrm{em}} \alpha_q
\, .
\end{align}
The mass of vector boson $m_V$ for photon is set to zero, \textit{i.e.}, $m_{\gamma} = 0~\mathrm{GeV}$, while for $Z$ boson $m_Z$ is set at $91.1876~\mathrm{GeV}$.
The fine structure constants $\alpha_{\mathrm{em}}$ used in phenomenological studies are $1 / 137$ and $1 / 128$ for $V = \gamma$ and $V = Z$, respectively, and $\alpha_s$ is the strong coupling constant.
The parameter $\alpha_q$ for $\gamma + h$ production is:
\begin{align}
\alpha_q \equiv e_q^2
\, ,
\end{align}
where $e_q$ denotes the electric charge of the quark.
For $Z+h$ production, $\alpha_q$ is given by:
\begin{align}
\alpha_q
\equiv
\frac{1 - 4 \abs{e_q} \sin[2](\theta_W) + 8 e_q^2 \sin[4](\theta_W)}{8 \sin[2](\theta_W) \cos[2](\theta_W)}
\, ,
\end{align}
where $\cos{\theta_W} = 0.88168$.
The collinear momentum fractions $x_a$ and $x_b$ are given by:
\begin{align}
x_a
& =
\frac{1}{\sqrt{s}}
\pqty{
\sqrt{p_{V,T}^2 + m_V^2} e^{y_V}
+
p_{V,T} e^{y_c}
}
\, , \label{e.x_a} \\
x_b
& =
\frac{1}{\sqrt{s}}
\pqty{
\sqrt{p_{V,T}^2 + m_V^2} e^{-y_V}
+
p_{V,T} e^{-y_c}
}
\, , \label{e.x_b}
\end{align}
where $\sqrt{s}$ is the center-of-mass energy for the collision.
Finally, $\hat{s}$, $\hat{t}$ and $\hat{u}$ appearing in \cref{e.hard-qqb,e.hard-qg} are the partonic Mandelstam variables.
To define them,
we first list the three relevant partonic processes:
\begin{align}
q \pqty{p_a} + \overline{q} \pqty{p_b}
\to
V \pqty{p_V} + g \pqty{p_c}
\, , \label{e.partonic-process-2} \\
q \pqty{p_a} + g \pqty{p_b}
\to
V \pqty{p_V} + q \pqty{p_c}
\, , \label{e.partonic-process-1}\\
\overline q \pqty{p_a} + g \pqty{p_b}
\to
V \pqty{p_V} + \overline q \pqty{p_c}
\, , \label{e.partonic-process-3}
\end{align}
where $q$, $\overline{q}$ and $g$ denote quark, antiquark and gluon, respectively, and $V$ denotes the vector boson.
The partonic Mandelstam variables are then defined as:
\begin{align}
\hat{s}
\equiv
\pqty{p_a + p_b}^2
\, , \quad
\hat{t}
\equiv
\pqty{p_a - p_V}^2
\, , \quad
\hat{u}
\equiv
\pqty{p_b - p_V}^2
\, . \label{e.partonic-Mandelstam-variables}
\end{align}

The NLO hard function used in this paper is given in \cite{Chien:2022wiq}.
Writing down the NLL resummation requires the hard anomalous dimension up to one-loop order \cite{Becher:2009th, Chien:2019gyf}:
\begin{align}
&
\gamma_{\mu}^H \pqty{\alpha_s}
=
-
\sum_{i = a, b, c}
\gamma_i \bqty{\alpha_s \pqty{\mu}}
\nonumber \\
& \quad +
\Gamma_{\mathrm{cusp}} \bqty{\alpha_s \pqty{\mu}}
\Bigg[
C_a \ln(\frac{\hat{u}^2}{p_{V,T}^2 \mu^2})
+
C_b \ln(\frac{\hat{t}^2}{p_{V,T}^2 \mu^2})
\nonumber \\
& \qquad \qquad \qquad \qquad +
C_c \ln(\frac{p_{V,T}^2}{\mu^2})
\Bigg]
+
\order{\alpha_s^2}
\, ,
\end{align}
where the cusp anomalous dimensions $\Gamma_{\mathrm{cusp}}$ and non-cusp anomalous dimensions $\gamma_i$ are expanded as:
\begin{align}
\Gamma_{\mathrm{cusp}} \bqty{\alpha_s(\mu)}
& =
\sum_{n = 1} \Gamma_{n-1} \pqty{\frac{\alpha_s}{4 \pi}}^n
\, , \label{e.cusp-anomalous-dimension} \\
\gamma_i \bqty{\alpha_s(\mu)}
& =
\sum_{n = 1} \gamma_{n-1}^i \pqty{\frac{\alpha_s}{4 \pi}}^n
\, . \label{e.non-cusp-anomalous-dimension}
\end{align}
For the expansion we keep the following terms \cite{Korchemsky:1987wg, Becher:2006mr, Jain:2011xz}:
\begin{align}
\Gamma_0
& =
4
\, , \quad
\gamma_0^q
=
6 C_F
\, , \quad
\gamma_0^g
=
2 \beta_0
\, , \\
\Gamma_1
& =
C_A\pqty{\frac{268}{9} - \frac{4 \pi^2}{3}} - \frac{80}{9} T_F n_f
\, ,
\end{align}
where $C_F = 4/3$, $C_A = 3$, $T_F = 1/2$ and $n_f$ is the number of flavors.
And $\beta_0$ is defined as:
\begin{equation}
\beta_0
\equiv
\frac{11}{3} C_A
-
\frac{4}{3} T_F n_f
\, .
\end{equation}

Next, we provide a brief overview of the TMD PDFs $f_{1,i/p}^{(u)} \pqty{x_i, b, \mu, \zeta_i/\nu^2}$ for an incoming parton $i$.
As discussed in \cite{Kang:2023oqj}, two scales other than $\mu$ are involved in the ``unsubtracted'' TMD PDFs $f_{1,i/p}^{(u)} \pqty{x_i, b, \mu, \zeta_i/\nu^2}$: the Collins-Soper scale $\zeta$ \cite{Collins:2011zzd, Boussarie:2023izj, Ebert:2019okf} and the rapidity renormalization scale $\nu$ \cite{Chiu:2012ir}.
For both quark and gluon TMD PDFs $f_{1,i/p}^{(u)}$, the rapidity divergences can be canceled by subtracting the square root of the standard soft function $S_{n_i \cdot \overline{n}_i} \pqty{b, L, L_{\nu}}$.
Denoted by $n_i$ is the light-like directional four-vector of an incoming or outgoing parton moving with a momentum $p_i$ defined in \cref{e.partonic-process-1,e.partonic-process-2,e.partonic-process-3}, and $L$ and $L_{\nu}$ will be defined in \cref{e.S_nnbar}.
Adopting dimensional regularization in $4 - 2 \epsilon$ space-time dimensions and the rapidity regulator $\eta$ discussed in \cite{Chiu:2012ir}, the standard soft function is given by \cite{Collins:2011zzd, Kang:2021ffh}:
\begin{align}
S_{n_i \cdot \overline{n}_i} \pqty{b, L, L_{\nu}}
& =
1 -
\frac{\alpha_s C_i}{2 \pi}
\bigg[2 \pqty{\frac{2}{\eta} + L_{\nu}} \pqty{\frac{1}{\epsilon} + L}
\nonumber \\
& \qquad \qquad \qquad -
\frac{2}{\epsilon^2} + L^2 + \frac{\pi^2}{6} \bigg]
, \label{e.S_nnbar}
\end{align}
where $L \equiv \ln(\mu^2 / \mu_b^2)$ with $\mu_b \equiv 2 e^{-\gamma_E} / b$, $L_{\nu}  \equiv \ln(\nu^2 / \mu^2)$, $C_i$ takes the value of $C_F$ or $C_A$ for quark and gluon TMD PDFs, respectively, and $\gamma_E$ is the Euler-Mascheroni constant.
Using $S_{n_i \cdot \overline{n}_i}$ from \cref{e.S_nnbar}, we additionally define the ``subtracted'' parton distribution $f_{1,i/p} \pqty{x_i, b, \mu, \zeta_i}$ that is free from the rapidity divergence \cite{Collins:2011zzd}:
\begin{align}
f_{1,i/p} \pqty{x_i, b, \mu, \zeta_i}
& =
f_{1,i/p}^{(u)} \pqty{x_i, b, \mu, \frac{\zeta_i}{\nu^2}}
\nonumber \\
& \quad \times
\sqrt{S_{n_i \cdot \overline{n}_i} \pqty{b, L, L_{\nu}}}
\, .
\end{align}

The TMD evolution for the ``subtracted'' TMD PDFs is now given by the Collins-Soper evolution and the renormalization group equation, each associated with the Collins-Soper scale $\zeta$ \cite{Collins:2011zzd, Boussarie:2023izj} and the scale $\mu$, respectively.
The evolution equations are given by:
\begin{align}
\frac{\dd{}}{\dd{\ln\sqrt{\zeta}}}
\ln{f_{1,i/p} \pqty{x_i, b, \mu, \zeta}}
& =
K (b, \mu)
\, , \\
\frac{\dd{}}{\dd{\ln{\mu}}}
\ln{f_{1,i/p} \pqty{x_i, b, \mu, \zeta_i}}
& =
\gamma_{\mu, i}^f \bqty{\alpha_s (\mu), \frac{\zeta_i}{\mu^2}}
\, ,
\end{align}
where $K \pqty{b, \mu}$ denotes the Collins-Soper evolution kernel \cite{Collins:2011zzd, Boussarie:2023izj, Moult:2022xzt, Duhr:2022yyp}, and $\gamma_{\mu, i}^f \bqty{\alpha_s (\mu), \zeta_i/\mu^2}$ is the $\mu$ evolution kernel that evolves $f_{1,i/p} \pqty{x_i, b, \mu, \zeta_i}$ in scale $\mu$ at fixed $\zeta = \zeta_i$.
Up to two-loop order, $\gamma_{\mu, i}^f$ is given by:
\begin{align}
\gamma_{\mu, i}^f \bqty{\alpha_s (\mu), \frac{\zeta_i}{\mu^2}}
=
& -
C_i \Gamma_{\mathrm{cusp}} \bqty{\alpha_s(\mu)}
\ln(\frac{\zeta_i}{\mu^2})
\nonumber \\
& +
\gamma_i \bqty{\alpha_s (\mu)}
\, ,
\end{align}
where $C_i = C_F$ $(C_A)$ for $i = q$ $(g)$, and $\Gamma_{\mathrm{cusp}}$ and $\gamma_i$ are the cusp and non-cusp anomalous dimensions, respectively.
They are given in \cref{e.cusp-anomalous-dimension,e.non-cusp-anomalous-dimension}.

Solving the renormalization group equations on $\zeta$ and $\mu$ while taking into account the non-perturbative contributions from the large $b \gg 1/\Lambda_{\mathrm{QCD}}$ region, we obtain the expressions for TMD PDFs:
\begin{align}
& f_{1,i/p} \pqty{x_i, b, \mu, \zeta_i}
=
f_{1,i/p} \pqty{x_i, b,\mu_{b_*},\mu_{b_*}^2}
\label{e.TMD_standard} \\
& \quad \times
\exp[-S_{\mathrm{NP}}^i (b, Q_0, \zeta_i)]
\exp[-S_{\mathrm{pert}}^i (\mu, \mu_{b_*}, \zeta_i)]
\nonumber \, ,
\end{align}
where we evolve the TMD PDFs from initial scales $\pqty{\mu_0, \zeta_0}$ to final scales $\pqty{\mu, \zeta_i = \hat{s}}$.
We have chosen the following initial scales $\mu_0 = \sqrt{\zeta_0} = \mu_{b_*}$.
As usual, we define $\mu_{b_*} \equiv 2 e^{-\gamma_E} / b_*$ and $b_* \equiv b / \sqrt{1 + b^2/b_{\max}^2}$ with $b_{\max} = 1.5~\mathrm{GeV}^{-1}$ following the $b_*$-prescription \cite{Echevarria:2020hpy, Sun:2014dqm, Isaacson:2023iui, Collins:1984kg}.
The choice of $\zeta_i = \hat{s}$ for the incoming partons will be discussed in \cref{ss.TEEC}.
The perturbative Sudakov factor $S_{\mathrm{pert}}^i \pqty{\mu, \mu_{b_*}, \zeta_i}$ is given by:
\begin{align}
S_{\mathrm{pert}}^i \pqty{\mu, \mu_{b_*}, \zeta_i}
& =
- K \pqty{b_*, \mu_{b_*}} \ln(\frac{\sqrt{\zeta_i}}{\mu_{b_*}})
\nonumber \\
& \quad -
\int_{\mu_{b_*}}^{\mu} \frac{\dd{\mu'}}{\mu'} \gamma_{\mu, i}^f \bqty{\alpha_s (\mu'), \frac{\zeta_i}{\mu^{\prime 2}}}
\, ,
\end{align}
and the non-perturbative Sudakov factor for $i = q$ is given by \cite{Sun:2014dqm, Echevarria:2020hpy}:
\begin{align}
S_{\mathrm{NP}}^i \pqty{b, Q_0, \zeta_i}
=
\frac{g_2}{2} \ln(\frac{b}{b_*}) \ln(\frac{\sqrt{\zeta_i}}{Q_0})
+
g_1^q b^2
\, , \label{e.S_NP-parameterization-for-TMD-PDFs}
\end{align}
with $Q_0^2 = 2.4~\mathrm{GeV}^2$, $g_2 = 0.84$ and $g_1^q = 0.106~\mathrm{GeV}^{2}$.
For $i = g$, we will follow \cite{Balazs:1997hv, Balazs:2000wv, Balazs:2007hr} and adopt the following parameterization:
\begin{align}
S_{\mathrm{NP}}^i \pqty{b, Q_0, \zeta_i}
=
\frac{C_A}{C_F} \frac{g_2}{2} \ln(\frac{b}{b_*}) \ln(\frac{\sqrt{\zeta_i}}{Q_0})
+
g_1^g b^2
\, ,
\end{align}
where we set $g_1^g = g_1^q$.
Throughout this paper, we work at the next-to-leading logarithmic (NLL) level in resummation accuracy, where $K \pqty{b_*, \mu_{b_*}} = 0$.

In the conventional TMD approach \cite{Boussarie:2023izj}, one can express $f_{1,i/p} \pqty{x, b, \mu_{b_*}, \mu_{b_*}^2}$ in terms of the collinear parton distribution functions through operator product expansion:
\begin{align}
f_{1,i/p} \pqty{x_i, \mu_{b_*}}
& =
\sum_j \int _{x_i}^1 \frac{\dd{x'}}{x'}
C_{i \leftarrow j} \pqty{\frac{x_i}{x'}, \mu_{b_*}} 
f_{j/p} \pqty{x', \mu_{b_*}}
\nonumber \\
& \equiv
\bqty{C_{i \leftarrow j} \otimes f_{j/p}} \pqty{x_i, \mu_{b_*}}
\, , \label{e.matching-TMD-PDFs}
\end{align}
where we adopt a shorthand notation $f_{1,i/p} \pqty{x_i, \mu_{b_*}}$ for TMD PDFs and $f_{j/p} \pqty{x, \mu}$ denotes the proton collinear parton distribution of flavor $j$.
The matching coefficients $C_{i \leftarrow j}$ are perturbatively calculable and can be found in \textit{e.g.}, \cite{Aybat:2011zv, Collins:2011zzd, Kang:2015msa, Echevarria:2016scs, Luo:2019szz, Echevarria:2020hpy, Luo:2020epw, Ebert:2020yqt}.

The nuclear modification is considered by substituting the nuclear collinear PDFs $f_{j/A}$ for the vacuum collinear PDFs $f_{j/p}$ in \cref{e.matching-TMD-PDFs}.
In this work we adopt the EPPS16 set \cite{Eskola:2016oht} for both gold (Au) and lead (Pb).
For the proton PDFs, we use CT14nlo \cite{Dulat:2015mca} to be consistent with the choice made in EPPS16.
Additionally, we follow the work in \cite{Alrashed:2021csd} and introduce a modified non-perturbative parameter $g_{1,A}^q$ in place of $g_1^q$ in \cref{e.S_NP-parameterization-for-TMD-PDFs}:
\begin{align}
g_{1,A}^q
=
g_1^q
+
a_N L
\, ,
\end{align}
where $L \equiv A^{1/3} - 1$ with $A$ being the mass number of the nucleus.
Our choice of nuclear broadening parameter $a_N = 0.016$ GeV$^2$ was determined in \cite{Alrashed:2021csd} from fitting experimental data using EPPS16 set \cite{Eskola:2016oht} as the collinear baseline.
The parameters $a_N$, as well as $b_N$ which will be discussed later in \cref{ss.TEEC}, are fitted from the semi-inclusive deep inelastic scattering (SIDIS) in electron-nucleus collisions and Drell-Yan (DY) dilepton production in proton-nucleus collisions \cite{Alrashed:2021csd}.
Although the universality of these parameters across SIDIS, DY and TEEC observables remains an assumption, future measurements of TEEC and other TMD observables will ultimately assess its validity.

\subsection{Soft function}
\label{ss.soft-function}

The soft function for $V+h$ production can be computed with the rapidity regulator \cite{Chiu:2011qc}, and the expressions are given by \cite{Buffing:2018ggv, Li:2020bub, Kang:2023oqj, Fang:2023thw}:
\begin{align}
S_{abc} \pqty{b, \mu, \nu}
=
1
& -
\boldsymbol{T}_a \cdot \boldsymbol{T}_b
S_{n \cdot \overline{n}}^{\pqty{1}} \pqty{b, L, L_{\nu} + \ln(n_{ab})}
\nonumber \\
& -
\boldsymbol{T}_b \cdot \boldsymbol{T}_c
S_{n \cdot \overline{n}}^{\pqty{1}} \pqty{b, L, L_{\nu} + \ln(n_{bc})}
\nonumber \\
& -
\boldsymbol{T}_c \cdot \boldsymbol{T}_a
S_{n \cdot \overline{n}}^{\pqty{1}} \pqty{b, L, L_{\nu} + \ln(n_{ca})}
\, ,
\end{align}
where $\boldsymbol{T}_{a,b,c}$ are color factors, and $S_{n \cdot \overline{n}}^{\pqty{1}}$ is the NLO component defined by the relation $S_{n_i \cdot \overline{n}_i} = 1 + C_{i} S_{n \cdot \overline{n}}^{\pqty{1}}$ for the standard soft function given in \cref{e.S_nnbar}.
We have also defined $n_{ab} \equiv n_a \cdot n_b / 2$.

Using color conservation $\boldsymbol{T}_a + \boldsymbol{T}_b + \boldsymbol{T}_c = 0$, we obtain:
\begin{align}
S_{abc} \pqty{b, \mu, \nu}
& =
1
+
\frac{1}{2}
\pqty{\boldsymbol{T}_a^2 + \boldsymbol{T}_b^2 + \boldsymbol{T}_c^2}
S_{n \cdot \overline{n}}^{\pqty{1}} \pqty{b, L, L_{\nu}}
\nonumber \\
& -
\frac{\alpha_s L}{2 \pi}
\sum_{\mathrm{cyc}} \pqty{\boldsymbol{T}_a^2 + \boldsymbol{T}_b^2 - \boldsymbol{T}_c^2}
\ln(n_{ab})
\nonumber \\
& -
\frac{\alpha_s L}{2 \pi}
\frac{1}{\epsilon}
\sum_{\mathrm{cyc}} \pqty{\boldsymbol{T}_a^2 + \boldsymbol{T}_b^2 - \boldsymbol{T}_c^2}
\ln(n_{ab})
\, ,
\end{align}
where $\boldsymbol{T}_{q(g)}^2 = C_F$ $(C_A)$ and $\sum_{\mathrm{cyc}}$ represents a cyclic summation over the indices $a$, $b$ and $c$.

Next we define the so-called ``proper" TMD global soft function which does not depend on the rapidity scale $\nu$:
\begin{align}
S_{abc}^{g,\mathrm{bare}} (b, \mu)
=
\frac{1}
{
\sqrt{
S_{n_a \cdot \overline{n}_a}
S_{n_b \cdot \overline{n}_b}
S_{n_c \cdot \overline{n}_c}
}
}
S_{abc} (b, \mu, \nu)
\, .
\end{align}
The renormalized TMD global soft function $S_{abc}^g (b, \mu)$ is:
\begin{align}
S_{abc}^g (b, \mu)
=
1
-
\frac{\alpha_s L}{2 \pi}
\sum_{\mathrm{cyc}} \pqty{\boldsymbol{T}_a^2 + \boldsymbol{T}_b^2 - \boldsymbol{T}_c^2}
\ln(n_{ab})
\, .
\end{align}
The renormalization group equation for $S_{abc}^g$ is given as:
\begin{align}
\frac{\dd{}}{\dd{\ln{\mu}}}
\ln{S_{abc}^g (b, \mu)}
=
\gamma_{\mu}^S (\alpha_s)
\, ,
\end{align}
where the anomalous dimension $\gamma_{\mu}^S$ is given by:
\begin{align}
\gamma_{\mu}^S (\alpha_s)
=
-\frac{\alpha_s}{\pi}
\sum_{\mathrm{cyc}}
\pqty{\boldsymbol{T}_a^2 + \boldsymbol{T}_b^2 - \boldsymbol{T}_c^2}
\ln(n_{ab})
\, .
\end{align}
One can work out the scalar products $n_{ab}$ in the above equation \cite{Gao:2023ivm}:
\begin{align}
n_{ab}
=
1
\, , \quad
n_{ac}
=
\frac{-\hat{u}}{\hat{s} - m_V^2}
\, , \quad
n_{bc}
=
\frac{-\hat{t}}{\hat{s} - m_V^2}
\, ,
\end{align}
where the partonic Mandelstam variables $\hat{s}$, $\hat{t}$ and $\hat{u}$ are given in \cref{e.partonic-Mandelstam-variables}.

\subsection{TEEC jet function}
\label{ss.TEEC}

We have introduced the ``unsubtracted'' TEEC jet function $J_c^{(u)} \pqty{b, \mu, \zeta_c/\nu^2}$ in \cref{e.unsubtracted-factorization}.
Its relation to the ``unsubtracted'' transverse momentum dependent fragmentation functions (TMD FFs) is given by \cite{Moult:2018jzp}:
\begin{align}
J_c^{(u)} \pqty{b,\mu,\frac{\zeta_c}{\nu^2}}
\equiv
\sum_h \int_0^1 \dd{z}
z D^{(u)}_{1, h/c} \pqty{z, b, \mu, \frac{\zeta_c}{\nu^2}}
\, .
\end{align}
Similarly to the TMD PDFs, we define the ``subtracted'' TMD FFs in such a way that the standard soft function $\sqrt{S_{n_c \cdot \overline{n}_c} \pqty{b, L, L_{\nu}}}$ cancels out the rapidity divergence:
\begin{align}
D_{1, h/c} \pqty{z, b, \mu, \zeta_c}
& =
D^{(u)}_{1, h/c} \pqty{z, b, \mu, \frac{\zeta_c}{\nu^2}}
\nonumber \\
& \quad \times
\sqrt{S_{n_c \cdot \overline{n}_c} \pqty{b, L, L_{\nu}}}
\, .
\end{align}
The evolution equation is given by:
\begin{align}
\frac{\dd{}}{\dd{\ln{\mu}}}
\ln{D_{1, h/c} \pqty{z, b, \mu, \zeta_c}}
& =
\gamma_{\mu, c}^J \bqty{\alpha_s (\mu), \frac{\zeta_c}{\mu^2}}
\, ,
\end{align}
where $\gamma_{\mu, c}^J \bqty{\alpha_s (\mu), \zeta_{c}/\mu^2}$ is the evolution kernel that governs the $\mu$ scale evolution of $D_{1, h/c} \pqty{z, b, \mu, \zeta_c}$ at fixed $\zeta = \zeta_c$.
Up to two-loop order, $\gamma_{\mu, c}^J$ is given by:
\begin{align}
\gamma_{\mu, c}^J \bqty{\alpha_s (\mu), \frac{\zeta_c}{\mu^2}}
=
& -
C_c \Gamma_{\mathrm{cusp}} \bqty{\alpha_s(\mu)}
\ln(\frac{\zeta_c}{\mu^2})
\nonumber \\
& +
\gamma_c \bqty{\alpha_s (\mu)}
\, .
\end{align}
The values of $\zeta_a$, $\zeta_b$ and $\zeta_c$ can be found by demanding RG consistency, \textit{i.e.}:
\begin{align}
\gamma_{\mu}^H
+
\gamma_{\mu, a}^f
+
\gamma_{\mu, b}^f
+
\gamma_{\mu}^S
+
\gamma_{\mu, c}^J
=
0
\, ,
\end{align}
for both $q \overline{q} \to  V + g$ and $q g \to  V + q$ channels.
We find that with the choice $\zeta_a = \zeta_b = \hat{s}$, and $\zeta_c = \pqty{\hat{s} - m_V^2}^2 / \hat{s}$, the RG consistency is satisfied for both channels.
Subsequently, the corresponding ``subtracted'' TEEC jet function $J_c \pqty{b, \mu, \zeta_c}$ can be written as:
\begin{align}
J_c \pqty{b, \mu, \zeta_c}
\equiv
\sum_h \int_0^1 \dd{z}
z D_{1, h/c} \pqty{z, b, \mu, \zeta_c}
\, .\label{e.TEEC-JFs}
\end{align}
The TMD FFs $D_{1,h/c} \pqty{z, b, \mu, \zeta_c}$ with QCD evolution are given by:
\begin{align}
D_{1,h/c} \pqty{z, b, \mu, \zeta_c}
& =
\sum_i \int_z^1 \frac{\dd{y}}{y}
\hat{C}_{i \leftarrow c} \pqty{\frac{z}{y}, b}
D_{h/i} \pqty{y,\mu_{b_*}}
\nonumber \\
& \quad \times
\exp[-S_{\mathrm{pert}} \pqty{\mu, \mu_{b_*}, \zeta_c}]
\nonumber \\
& \quad \times
\exp[-S_{\mathrm{NP}}^c \pqty{z, b, Q_0, \zeta_c}]
\, , \label{e.TMD-FFs}
\end{align}
where $Q_0^2 = 2.4~\mathrm{GeV}^2$, $D_{h/i}$ are the vacuum collinear fragmentation functions (FFs) and the matching coefficients $\hat{C}_{i\leftarrow c}$ can be found in \cite{Echevarria:2020hpy, Luo:2019szz, Luo:2020epw, Ebert:2020yqt}.
The corresponding non-perturbative Sudakov factor $S_{\mathrm{NP}}^c \pqty{z, b, Q_0, \zeta_c}$ for quark and gluon are respectively given by \cite{Kang:2017glf}:
\begin{align}
S_{\mathrm{NP}}^q \pqty{z, b}
& =
\frac{g_2}{2} \ln(\frac{b}{b_*}) \ln(\frac{\sqrt{\zeta_c}}{Q_0}) + g_1^D \frac{b^2}{z^2}
\, , \label{e.quark-S_NP-parameterization-for-TEEC-JFs} \\
S_{\mathrm{NP}}^g \pqty{z, b}
& =
\frac{C_A}{C_F}\frac{g_2}{2} \ln(\frac{b}{b_*}) \ln(\frac{\sqrt{\zeta_c}}{Q_0}) + g_1^D \frac{b^2}{z^2}
\, , \label{e.gluon-S_NP-parameterization-for-TEEC-JFs}
\end{align}
with $g_2 = 0.84$ and $g_1^D = 0.042~\mathrm{GeV}^2$ \cite{Sun:2014dqm, Echevarria:2020hpy} and we have omitted the $Q_0$ and $\zeta_c$ dependence for brevity.

Following the procedures in \cite{Kang:2023oqj},  we use the LO matching coefficients $\hat{C}_{i \leftarrow c} \pqty{z,b} = \delta_{ic} \delta \pqty{1-z}$ in \cref{e.TMD-FFs} and fit the $z$-integration in \cref{e.TEEC-JFs} with \cite{Li:2021txc, Kang:2023oqj}:
\begin{align}
& \sum_h \int_0^1 \dd{z}
z D_{h/i} \pqty{z, \mu_{b_*}}
\exp(-g_1^D \frac{b^2}{z^2})
\nonumber \\
& \equiv
\exp[-S_{\mathrm{NP}}^{\mathrm{TEEC}} (b)]
\, , \label{e.S_NP-TEEC}
\end{align}
with the functional form:
\begin{align}
S_{\mathrm{NP}}^{\mathrm{TEEC}} (b)
=
N b^{\alpha}
\pqty{1 + r b^{\beta}}
\, . \label{e.fitted-S_NP-TEEC}
\end{align}
Here $N$, $\alpha$, $r$ and $\beta$ are the fit parameters and their extracted values are given in \cref{t.TEEC-parameters}.
We emphasize that the conditions $\alpha > 0$ and $\beta > 0$ are imposed to guarantee that the second term ($N r b^{\alpha + \beta}$) maintains a strictly larger exponent than the first term ($N b^{\alpha}$).
The parameterization chosen here is different from the previous choice made in \cite{Kang:2023oqj}, since the new parameterization is more flexible and can lead to a successful fit in both vacuum and nuclear environment.
As for the choice of collinear FFs that enter \cref{e.S_NP-TEEC}, we use the 2021 DSS parameterization \cite{Borsa:2021ran} for neutral and charged pions, \textit{i.e.}, $\pi^0$ and $\pi^{\pm}$.
In order to remove the bias from including only pion FFs, we normalize the left-hand side of \cref{e.S_NP-TEEC} by $\sum_{\pi^0, \pi^{\pm}} \int_0^1 \dd{z} z D_{h/i} \pqty{z, \mu_{b_*}}$.
Note that in \cref{e.fitted-S_NP-TEEC}, the same set of parameters in $S_{\mathrm{NP}}^{\mathrm{TEEC}} (b)$ is used for quark and gluon since the sum rule $\sum_h \int_0^1 \dd{z} z D_{h/i} (z, \mu_{b_*}) = 1$ holds for $i = q$ and $g$.
Additionally, although the FFs for charged and neutral hadrons are currently available in vacuum \cite{Borsa:2023zxk}, they have not been fitted in nuclear environment yet.
We therefore choose to use pion FFs in vacuum, so as to be consistent with the LIKEn pion FFs \cite{Zurita:2021kli} we used in nuclear environment.
We then have the final expressions for the quark and gluon TEEC jet functions:
\begin{align}
&
J_q \pqty{b, \mu, \zeta_c}
=
\exp[-S_{\mathrm{pert}} \pqty{\mu, \mu_{b_*}, \zeta_c}]
\\
& \qquad \times
\exp[-\frac{g_2}{2} \ln(\frac{b}{b_*}) \ln(\frac{\sqrt{\zeta_c}}{Q_0}) - S_{\mathrm{NP}}^{\mathrm{TEEC}} (b)]
\nonumber \, , \\
&
J_g \pqty{b, \mu, \zeta_c}
=
\exp[-S_{\mathrm{pert}} \pqty{\mu, \mu_{b_*}, \zeta_c}]
\\
& \qquad \times
\exp[-\frac{C_A}{C_F} \frac{g_2}{2} \ln(\frac{b}{b_*}) \ln(\frac{\sqrt{\zeta_c}}{Q_0}) - S_{\mathrm{NP}}^{\mathrm{TEEC}} (b)]
\nonumber \, .
\end{align}
As one can expect from the TMD factorization formalism of TMD FFs, the quark and gluon TEEC jet functions are only different by a color factor $C_A/C_F$ in the non-perturbative Sudakov factor.

The nuclear modification for TEEC jet function is considered following a procedure similar to the one for the TMD PDFs in \cref{ss.hard-function-and-TMD-PDFs}.
First we replace the vacuum collinear FFs $D_{h/i}$ in \cref{e.TMD-FFs} with the nuclear FFs $D_{h/i}^A$.
In this work, we use the LIKEn 2021 set \cite{Zurita:2021kli} for both gold (Au) and lead (Pb).
The fitted parameters are summarized in \cref{t.TEEC-parameters}.
Additionally, we follow the work in \cite{Alrashed:2021csd} and introduce a new parameter $g_{1,A}^D$ in place of $g_1^D$ in \cref{e.quark-S_NP-parameterization-for-TEEC-JFs,e.gluon-S_NP-parameterization-for-TEEC-JFs} such that:
\begin{align}
g_{1,A}^D
=
g_1^D
+
b_N L
\, ,
\end{align}
where $L \equiv A^{1/3} - 1$ with $A$ being the mass number of the nucleus, and $b_N = 0.0097$ GeV$^2$ has been determined from fitting experimental data~\cite{Alrashed:2021csd}. See also Ref.~\cite{Alrashed:2023xsv}.

\begin{table}[!h]
\setcellgapes{2.5 pt}
\makegapedcells
\begin{tabular}{|c|c|c|c|c|}
\hline
nucleus & $N \bqty{\mathrm{GeV}^{\alpha}}$ & $\alpha$ & $r \bqty{\mathrm{GeV}^{\beta}}$ & $\beta$ \\
\hline
$p$ & 0.45 & 0.42 & 0.63 & 1.00 \\
\hline
Au & $1.21 \pm 0.02$ & $0.56 \pm 0.01$ & $0.138 \pm 0.008$ & $1.21 \pm 0.02$\\
\hline
Pb & $1.22 \pm 0.03$ & $0.56 \pm 0.01$ & $0.136 \pm 0.007$ & $1.22 \pm 0.02$ \\
\hline
\end{tabular}
\caption{
The values of fitted parameters from \cref{e.fitted-S_NP-TEEC}.
The values are approximated to the second digits after decimal point.
The average values and standard deviation for Au and Pb are calculated from 29 fits corresponding to the Hessian sets given in LIKEn 2021 FFs \cite{Zurita:2021kli}.
The $r$ parameters for Au and Pb are displayed with three decimal places because their variations are smaller.
}
\label{t.TEEC-parameters}
\end{table}

An interesting point that we notice in \cref{t.TEEC-parameters} is that the fitted parameters for Au and Pb are very close to each other.
This is due to the fact that the mass numbers of Au and Pb are very similar (197 and 208 respectively), and the only distinction in the parameterization of nuclear effects (in both collinear and TMD FFs) for Au and Pb lies in their mass numbers.

Finally, one can write the factorization formula in \cref{e.unsubtracted-factorization} using the ``subtracted'' TMD PDFs and TEEC jet functions as:
\begin{widetext}
\begin{align}
\frac{\dd{\Sigma}}{\dd{\tau} \dd{y_V} \dd{p_{V,T}}}
& =
\sum_{a,b,c} \frac{p_{V,T}}{\sqrt{\tau}}
\int_0^{\infty} \frac{\dd{b}}{\pi}
\cos(2 b \sqrt{\tau} p_{V,T})
\int \dd{y_c}
H_{ab \to Vc} \pqty{y_V, p_{V,T}, m_V, \mu}
\nonumber \\
& \qquad \qquad \qquad \qquad \quad \times
x_a \, f_{1,a/p} \pqty{x_a, b, \mu, \zeta_a} \,
x_b \, f_{1,b/p} \pqty{x_b, b, \mu, \zeta_b}
J_c \pqty{b, \mu, \zeta_c}
S_{abc}^g \pqty{b, \mu}
\, . \label{e.factorization}
\end{align}
\end{widetext}
Following \cite{Alrashed:2021csd}, we assume that the nuclear TMD factorization in $pA$ collisions retains the same structure as the TMD factorization in $pp$ collisions, except that the free TMD distributions (including both TMD PDFs and TMD FFs) are replaced by the nuclear TMD distributions discussed earlier.
This assumption on nuclear TMD factorization has been examined in \cite{Alrashed:2021csd} using existing global data, including SIDIS in electron-nucleus collisions and DY dilepton production in proton-nucleus collisions.
We believe that the future measurements of the TEEC observable, as presented in this work, will provide a valuable contribution to further validating the assumption.

In performing phenomenological studies, we choose to evolve the hard function, TMD PDFs, TEEC jet function as well as the soft function to a common scale $\mu = \sqrt{m_V^2 + p_{V,T}^2}$ with $m_V = 91.1876~\mathrm{GeV}$ for $Z + h$ and $m_V = 0~\mathrm{GeV}$ for $\gamma + h$.
More details on the numerical values of relevant parameters will be discussed in \cref{s.phenomenology}.

\section{Phenomenology}
\label{s.phenomenology}

In this section, we provide numerical predictions for the TEEC using the factorization formula in \cref{e.factorization}, both with the RHIC and LHC kinematics.
Besides $pp$ collisions, $p\mathrm{Au}$ collisions at RHIC and $p\mathrm{Pb}$ collisions at LHC kinematics are also studied to assess nuclear TMD effects.

\begin{figure}[t]
\centering
\includegraphics[width = 0.47 \textwidth]{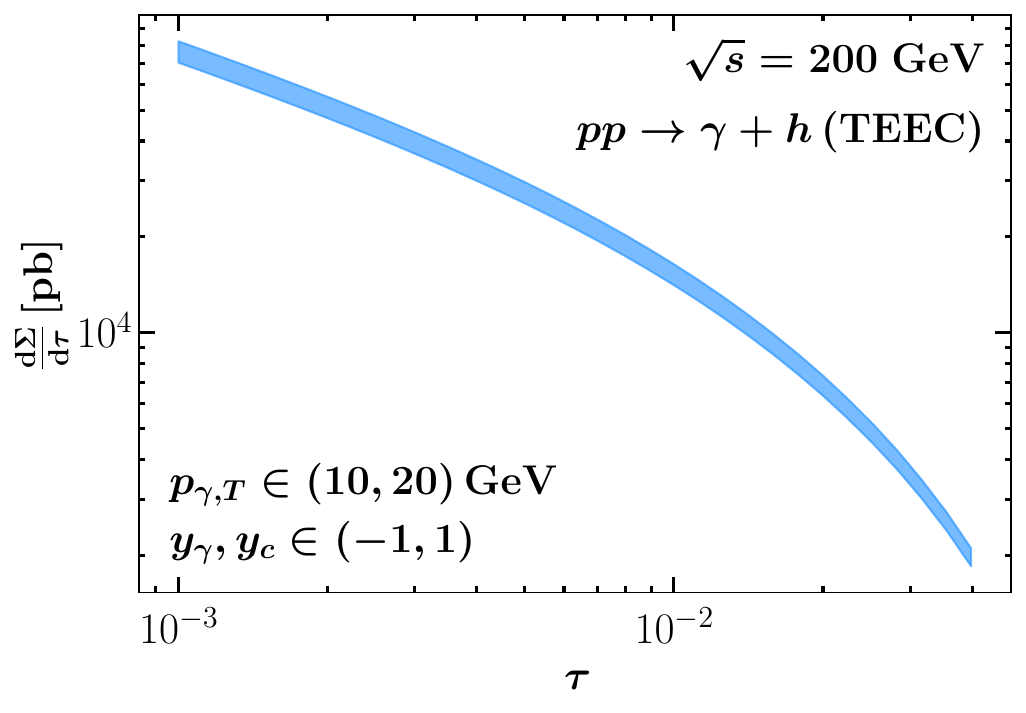}
\caption{
The predictions for the TEEC observable in $pp \to \gamma + h$ at sPHENIX kinematics.
Also shown in the figure are the kinematic requirements for photons.
The uncertainty band reflects variations of $\mu$ around its nominal value by factors of 0.5, 1, and 2, and represents the envelope of the resulting variations.
}
\label{f.TEEC-pp-sPHENIX}
\end{figure}

First, for the $pp$ collisions at RHIC, we choose $\sqrt{s} = 200~\mathrm{GeV}$, final-state photon $p_{V,T} \in \pqty{10, 20}~\mathrm{GeV}$ and rapidity range $y_V \in \pqty{-1, 1}$, $y_c \in \pqty{-1, 1}$.
This corresponds to the kinematics and detector acceptance at sPHENIX experiment.
The predictions for the TEEC in the process $pp \to \gamma + h$ at sPHENIX kinematics are shown in \cref{f.TEEC-pp-sPHENIX}.
The cross section decreases with increasing $\tau$, which aligns with expectation, as the final-state photon and hadrons deviate further from the back-to-back configuration as $\tau$ becomes larger, resulting in a lower event count.
In \cref{f.TEEC-pp-sPHENIX}, we also present the uncertainty band, which is derived by varying $\mu$ around its nominal value, $\mu = \sqrt{m_V^2 + p_{V,T}^2}$, by factors of 0.5, 1, and 2, and then taking the envelope of the resulting variations.

\begin{figure}[!h]
\centering
\includegraphics[width = 0.47 \textwidth]{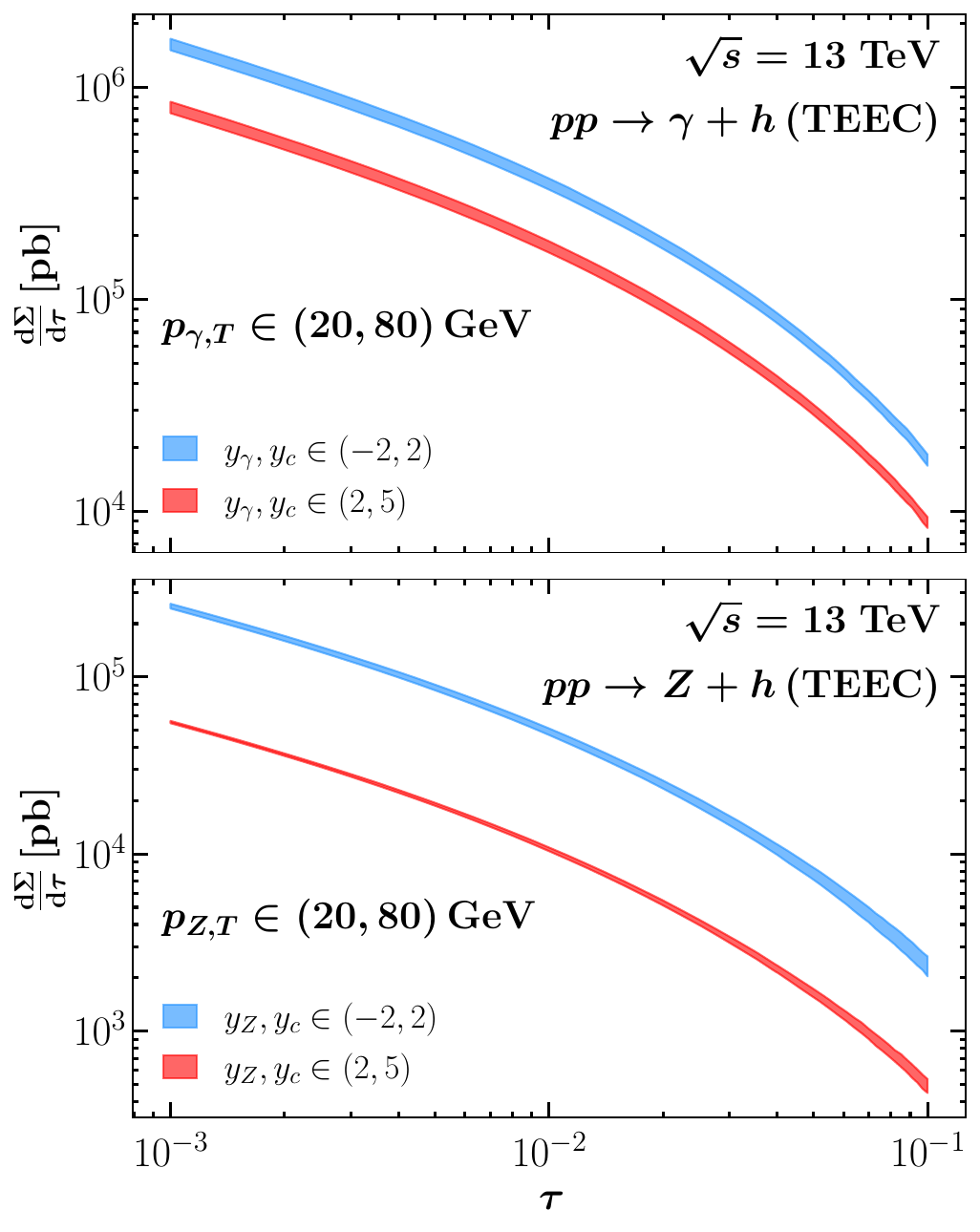}
\caption{
The predictions for the TEEC observable at mid and forward rapidity.
In the upper panel, TEEC observable in $pp \to \gamma + h$ at mid (blue) and forward (red) rapidity regions are shown, while in the lower panel, TEEC observable in $pp \to Z + h$ at mid (blue) and forward (red) rapidity regions are presented.
The uncertainty bands reflect variations of $\mu$ around its nominal value by factors of 0.5, 1, and 2, with the envelope of these variations defining the bands.
}
\label{f.TEEC-pp-CERN}
\end{figure}

In the upper panel of \cref{f.TEEC-pp-CERN}, we present the same observable in $pp \to \gamma + h$ at LHC energy.
We use two rapidity ranges, one with $y_{V,c} \in \pqty{-2, 2}$ corresponding to the mid-rapidity, the other one with $y_{V,c} \in \pqty{2, 5}$ corresponding to the forward rapidity region.
Finally in the lower panel of \cref{f.TEEC-pp-CERN}, we present the predicted $pp \to Z + h$ at LHC energy.
A similar trend that was predicted for RHIC is observed here, which again agrees with our expectation.
Additionally, the number of events at forward rapidity bin is less than the one at mid-rapidity bin, as expected.
Again, the uncertainty bands in \cref{f.TEEC-pp-CERN} reflect variations of $\mu$ around its nominal value by factors of 0.5, 1, and 2, with the envelope of the resulting variations defining the bands.

In order to study the nuclear effects, we present the result for $p\mathrm{Au}$ collision at sPHENIX at RHIC and $p\mathrm{Pb}$ collision at LHC.
First we define the nuclear modification factor as:
\begin{align}
R_{pA}
\equiv
\frac{1}{A}\frac{\dd{\Sigma^{pA}}}{\dd{\tau} \dd{y_V} \dd{p_{V,T}}}
\bigg /
\frac{\dd{\Sigma^{pp}}}{\dd{\tau} \dd{y_V} \dd{p_{V,T}}}
\, ,
\end{align}
where $A$ in the subscript represents the nuclear target.

\begin{figure}[h]
\centering
\includegraphics[width = 0.47 \textwidth]{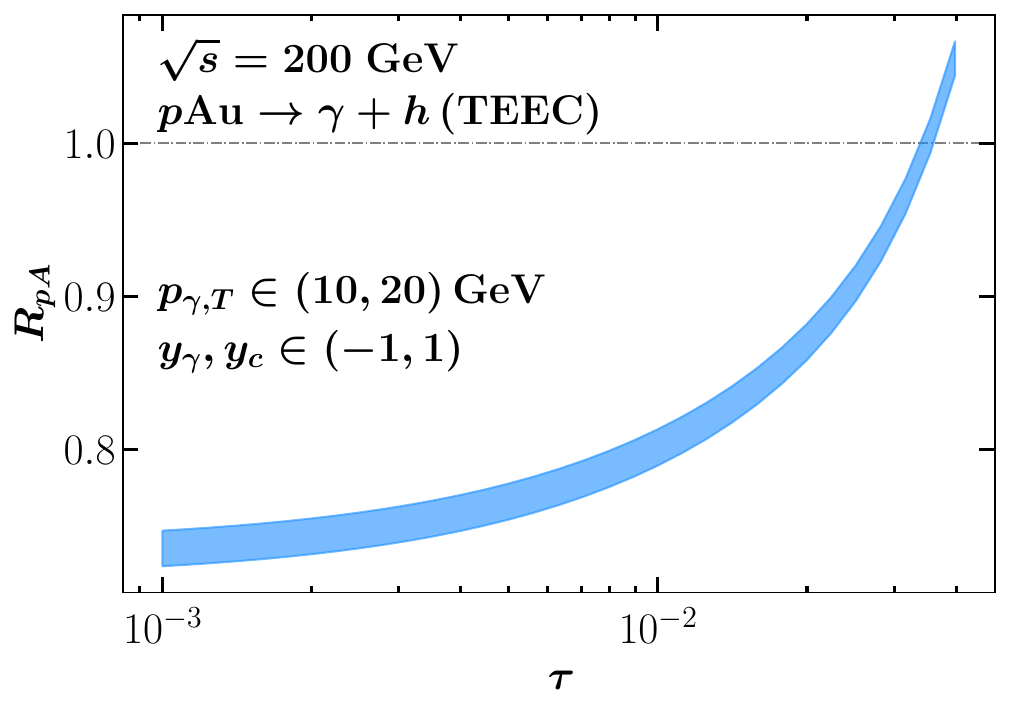}
\caption{
Nuclear modification for $p\mathrm{Au} \to \gamma + h$ at sPHENIX kinematics.
The kinematics are labeled in the plots.
The band represents the uncertainty from different members of nuclear collinear FFs \cite{Zurita:2021kli}.
}
\label{f.R_pA-sPHENIX}
\end{figure}

For $p\mathrm{Au}$ collision at RHIC, we plot the nuclear modification factor $R_{pA}$ at $\sqrt{s} = 200~\mathrm{GeV}$, with $p_T \in \pqty{10, 20}~\mathrm{GeV}$ and $y_{\gamma}, y_c \in \pqty{-1, 1}$.
This is shown in \cref{f.R_pA-sPHENIX}.
The band represents the uncertainties arising from variations in the choice of members from the nuclear collinear FFs \cite{Zurita:2021kli}.
At small $\tau$, a nuclear modification of $\sim 30\%$ can be expected.
On the other hand, the nuclear modification approaches or even exceeds 1 as $\tau$ gets larger.
This is due to nuclear modification manifesting itself as a broadening effect in the transverse momentum distribution.
To be more specific, larger $\tau$ values correspond to larger transverse momentum imbalance between the final-state boson and hadron, and since the transverse momentum gets smeared to larger values in the nuclear environment, the nuclear modification behaves as a suppression effect at smaller $\tau$ and enhancement at larger $\tau$.

\begin{figure}[h]
\centering
\includegraphics[width = 0.47 \textwidth]{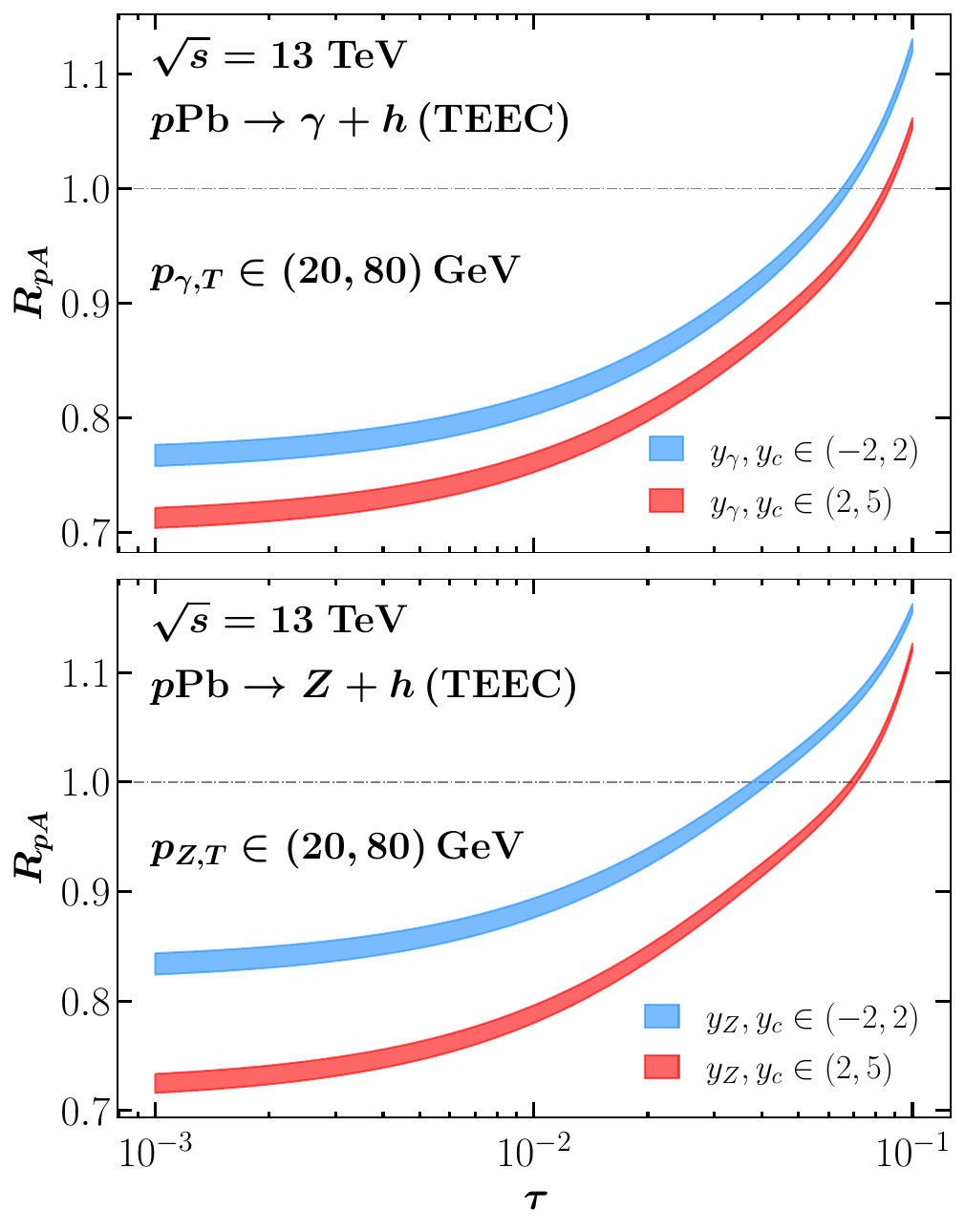}
\caption{
In the upper panel, nuclear modifications for $p\mathrm{Pb} \to \gamma + h$ at mid (blue) and forward (blue) rapidity are shown.
In the lower panel, nuclear modifications for $p\mathrm{Pb} \to Z + h$ at mid (blue) and forward (blue) rapidity regions are shown.
The kinematic requirements are summarized in the figure.
The bands represent the uncertainties arising from different choices of members from nuclear collinear FFs \cite{Zurita:2021kli}.
}
\label{f.R_pA-CERN}
\end{figure}

We have also plotted the nuclear modification predictions for the LHC at mid and forward rapidity in \cref{f.R_pA-CERN}. Similar behavior is observed between the RHIC and LHC kinematics.
Additionally, when comparing the $R_{pA}$ in the central and forward regions, we find that the $R_{pA}$ values are smaller in the forward rapidity region.
This occurs because, in the forward rapidity region, the value of $x_b$ (as given in \cref{e.x_b}) becomes very small, entering deeper into the shadowing region of nuclear modification.
This analysis highlights that the TEEC in $pA$ collisions serves as an excellent observable for studying nuclear modification effects on TMDs.
Specifically, the ability to measure a broad range of $\tau$ enables us to study the transverse momentum broadening in the nuclear medium, as encoded in the shape of the $\tau$ distribution.
Furthermore, by examining the $\tau$ distribution across different rapidity regions, we can investigate nuclear modification in the collinear parton distribution functions.


\section{Conclusions}
\label{s.conclusion}

In this paper, we have explored the transverse energy-energy correlator in the $pp$ collision for back-to-back $\gamma$-hadron and $Z$-hadron production at RHIC and the LHC.
In the transverse plane, the azimuthal angle difference $\phi$ between the final-state vector boson and the hadron is measured, and we provide a factorization formula for this event-shape observable as $\phi \to \pi$.
We present numerical results for TEEC in $pp$ collisions, and find that the TEEC cross section decreases as $\tau$ get larger.

Additionally, we study the nuclear modification in $p\mathrm{Au}$ collision at RHIC and $p\mathrm{Pb}$ collision at LHC.
We find that the nuclear modification factor can be $\sim 70\%$ at low $\tau$ region and grows with $\tau$.
This observation agrees with our expectation that the nuclear medium broadens the transverse momentum distribution and thus smears the cross section to larger $\tau$ values, or equivalently gives rise to greater transverse momentum imbalance. Furthermore, we investigate the nuclear modification of the TEEC at both central and forward rapidity regions. Our findings indicate that measurements at different rapidities offer valuable insights into the nuclear modification of the collinear component of the TMD in nuclei. 
This highlights that the TEEC observable can serve as an effective probe for studying nuclear modification in TMD physics.
Finally, the future measurement of the TEEC observables in proton-nucleus collisions could provide further validation of the assumptions made in the nuclear TMD factorization.

In summary, exploring the transverse energy-energy correlator in vector boson plus hadron production in proton-proton and proton-nucleus collisions presents a fertile ground for studying TMD physics, both in vacuum and nuclear environment.
We anticipate the insights obtained from TEEC observables will be crucial in enhancing our understanding of the fundamental aspects of strong interaction physics.
We encourage experiments at both RHIC and the LHC to carry out these measurements.

\section*{Acknowledgments}

Z.K. and J.P. are supported by the National Science Foundation under grant No.~PHY-1945471. S.L. is supported by U.S. Department of Energy under Contract No. DE-FG02-96ER40982.
F.Z. is supported by U.S. Department of Energy, Office of Science, Office of Nuclear Physics under grant Contract Number DESC0011090 and U.S. Department of Energy, Office of Science, National Quantum Information Science Research Centers, Co-design Center for Quantum Advantage (C2QA) under Contract No. DESC0012704.
Y.Z. is supported by the European Union ``Next Generation EU'' program through the Italian PRIN 2022 grant No. 20225ZHA7W.
Y.Z. is also supported by the Guangdong Major Project of Basic and Applied Basic Research No. 2020B0301030008, and the National Natural Science Foundation of China under Grants No. 12022512 and No. 12035007.
This work is also supported by the U.S. Department of Energy, Office of Science, Office of Nuclear Physics, within the framework of the Saturated Glue (SURGE) Topical Theory Collaboration.

\bibliographystyle{JHEP-2modlong.bst}
\bibliography{main.bib}
\end{document}